\documentclass[superscriptaddress, twocolumn, notitlepage, amsmath, amssymb, aps, prl, 10pt, letterpaper, showpacs]{revtex4-1}

\usepackage{graphicx,times,color,hyperref,tikz,soul}

\begin{document}

\title{Passive $\mathcal{PT}$-symmetric couplers without complex optical potentials}

\author{Yi-Chan Lee}
\affiliation{Physics Department, National Tsing-Hua University, Hsinchu 300, Taiwan}

\author{Jibing Liu}
\affiliation{Institute of Photonics Technologies, National Tsing-Hua University, Hsinchu 300, Taiwan}
\affiliation{College of Physics and Electronic Science, Hubei Normal University, Huangshi 435002, China}

\author{You-Lin Chuang}
\affiliation{Institute of Photonics Technologies, National Tsing-Hua University, Hsinchu 300, Taiwan}

\author{Min-Hsiu Hsieh}
\affiliation{Centre for Quantum Computation \& Intelligent Systems, Faculty of Engineering and Information Technology, University of Technology, Sydney, NSW 2007, Australia}

\author{Ray-Kuang Lee}
\affiliation{Physics Department, National Tsing-Hua University, Hsinchu 300, Taiwan}
\affiliation{Institute of Photonics Technologies, National Tsing-Hua University, Hsinchu 300, Taiwan}
\affiliation{Physics Division, National Center for Theoretical Science, Hsinchu 300, Taiwan}

 \email{$^{*}$rklee@ee.nthu.edu.tw}
\begin{abstract}
In addition to the implementation of parity-time ($\mathcal{PT}$)-symmetric optical systems by carefully and actively controlling the gain and loss, we show that a $2\times 2$ $\mathcal{PT}$-symmetric Hamiltonian has a unitarily equivalent representation without complex optical potentials in the resulting optical coupler. 
Through the Naimark dilation in operator algebra, passive $\mathcal{PT}$-symmetric couplers can thus be implemented with a refractive index of real values and asymmetric coupling coefficients. 
This opens up the possibility to implement general $\mathcal{PT}$-symmetric systems  with state-of-the-art asymmetric slab waveguides, dissimilar optical fibers, or cavities with chiral mirrors.
\end{abstract}

\pacs{42.79.Gn, 42.25.Bs, 11.30.Er, 78.20.Ci}

\maketitle

With spatial reflection and time reversal,  parity-time ($\mathcal{PT}$)-symmetric systems that could exhibit  entirely real and positive eigenvalue spectra have attracted considerable attention \cite{bender1998,bender2007}.  In {Bender and Boettcher}'s original proposal,  such a class of non-Hermitian systems reveals the possibility of removing the restriction of Hamiltonians from Hermiticity to a weaker $\mathcal{PT}$-symmetry. 
Nevertheless, it was pointed out that the no-signalling principle will be violated when applying the local $\mathcal{PT}$-symmetric operation on one of the entangled particles~\cite{YiChan}. Although situations become much more complicated when quantum entanglement is involved, $\mathcal{PT}$-symmetry could still be used as an interesting model for open systems in classical limit~\cite{PT-device}.

Based on the equivalence between the Schr\"odinger equation and the optical wave equation, 
classical optical systems have proven to be an excellent testbed for studying properties of $\mathcal{PT}$-symmetric systems. 
To satisfy $\mathcal{PT}$-symmetry,  optical systems with a complex potential   $U(x)=U^\ast (-x)$ are required for  one-dimensional (1D) optical couplers (planar waveguides or cavities)~\cite{klaiman2008}. 
As a result, the real part of this complex potential  is  an even function of the coordinate variable $x$; while the corresponding imaginary part is an odd one.
Several unique features, such as nonlinear soliton dynamics~\cite{soliton}, power oscillations in synthetic optical lattices~\cite{makris2008}, unidirectional invisibility~\cite{nature}, and loss-induced suppression of lasing~\cite{loss}, have been demonstrated on $\mathcal{PT}$-symmetric optical systems. 

Experimental demonstration of $\mathcal{PT}$-symmetry in optics has been realized with spatially balanced gain and loss of energy in a planar slab waveguide~\cite{Guo, Ruter}. 
However, it still remains challenging to keep gain and loss constantly balanced in optical devices.  In addition to active $\mathcal{PT}$-symmetric optical systems, passive $\mathcal{PT}$symmetry breaking has been experimentally demonstrated by externally modulating meta-materials on the Si-on-insulator platform~\cite{passive}. 
More recently, it was also revealed that a supersymmetric transformation can provide a versatile platform by synthesizing the refractive index profile in optical systems with $\mathcal{PT}$-symmetry~\cite{super}. 

In this Letter, we  demonstrate that any $2\times 2$ $\mathcal{PT}$-symmetric system can be unitarily transformed into another $\mathcal{PT}$-symmetric system with only real-valued Hamiltonians. This unitary transformation is constructed by applying the  Naimark dilation to embed a $\mathcal{PT}$-Hamiltonian in larger system dimensions~\cite{Naimark}. As a result, we show that the refractive index of real values and asymmetric coupling  coefficients of $\mathcal{PT}$-symmetric optical couplers are sufficient to satisfy the  $\mathcal{PT}$-symmetry condition. Our result has an immediate experimental implication since  asymmetric couplers in the slab waveguides have been well-studied~\cite{Marcuse}. With current optical device technologies, these asymmetric couplers  can be easily implemented with dissimilar optical fibers~\cite{Marcuse-85}, or through  unequal amplitudes at the two boundaries  in the two cladding layers~\cite{Snyder}. Furthermore, the asymmetric transmission of circularly polarzied wave is realized by applying metamaterial~\cite{Plum-2011}, where the metamaterial has been used to realize gain/loss $\mathcal{PT}$-symmetric system~\cite{Lawerance-2014}, showing that asymmetric coupling system is promising. Our proposal hence provides a more accessible platform  to study $\mathcal{PT}$-symmetric systems in classical optics, and overcomes the current difficulty of implementing balanced gain/loss  $\mathcal{PT}$-symmetric optical couplers. 

\begin{figure} [th]
	\centering
\includegraphics[width=8.0cm]{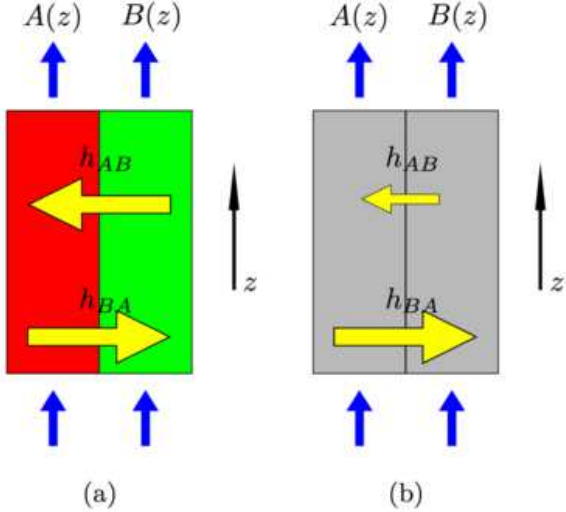}
		\caption{(Color online) (a) Illustration of a $\mathcal{PT}$-symmetric coupler with complex optical potentials for the Hamiltonian shown in Eq.~(4). Here, two eigenmodes for  the left and right channels in the paraxial wave approximation are denoted as $A(z)$ and $B(z)$, respectively. The channel with gain and loss are represented in color red (dark gray) and green (light gray). The magnitudes of gain and loss are the same to satisfy the required  $\mathcal{PT}$ symmetry; while the coupling strengths between two channels are the same, i.e., $h_{AB} = h_{BA}$.  
(b) Illustration of a $\mathcal{PT}$-symmetric coupler {\it without} complex optical potentials for the Hamiltonian shown in Eq.~(17a).  Asymmetry coupling strengths are depicted in different sizes of arrow, i.e., $h_{AB} \neq h_{BA}$, and gray (whole area) is used to represent where there is no gain and loss in each channel.}
\end{figure}

Let us consider a 1D optical coupler, as illustrated in Fig. 1, with  two eigenmodes  denoted as $A(z)$ and $B(z)$ for  the left and right channels in the paraxial wave approximation, respectively. The dynamics for this optical coupler satisfies a  Schr\"odinger-like wave equation:
\begin{eqnarray}
	i\, \frac{\partial \psi}{\partial z} = H\, \psi,
\end{eqnarray}
with $\psi = \left( A, B\right)^T$ and
\begin{eqnarray}
	H
	= s
	\left(
		  \begin{array}{cc}
		  	h_A 	& h_{AB}\\
			h_{BA}	& h_B
		  \end{array}
	\right),
\end{eqnarray}
where $s$ is the scaling factor of the system; $h_A$ and $h_B$ are the corresponding potentials (propagation constants) of the left channel and right channels, and $h_{AB}$ and $h_{BA}$ are the coupling strengths between two channels, respectively.  The optical Hamiltonian $H$ is $\mathcal{PT}$-symmetric when it commutes with the  $\mathcal{PT}$ operator,  i.e.,
\begin{align}
	[\mathcal{PT}, H] = 0,
\end{align}
where $\mathcal{P}$ is the spatial reflection operator that takes $x\rightarrow -x$, and $\mathcal{T}$ is the time reversal operator that takes $i\rightarrow -i$. 
It is easy to see that the eigenvalues of $H$ are always real when the eigenstates of a $\mathcal{PT}$-symmetric Hamiltonian are also the eigenstates of $\mathcal{PT}$. 
However, when the eigenstates of $H$ are no longer the eigenstates of $\mathcal{PT}$, the eigenvalues become complex. This is called spontaneous $\mathcal{PT}$ symmetry-breaking.

In the following, we will show how to construct a general  $\mathcal{PT}$-symmetric Hamiltonian  with only the real values of the refractive index involved. In the case of Hermitian Hamiltonians, the situation is much simpler since one can directly apply a unitary transformation to construct a real-valued Hamiltonian. 
In contrast, it is not clear if this is also true in the $\mathcal{PT}$-symmetric system, in which the inner product is defined differently from a Hermitian one. In the following, we answer this question positively.

Based on the Naimark dilation~\cite{Naimark}, the $\mathcal{PT}$-symmetric Hamiltonian is expanded into a Hermitian one with a larger system dimension. Consider the following $\mathcal{PT}$-symmetric Hamiltonian in Ref.~\cite{Ruter}: 
\begin{align}
	H
	= s 
	\left(
		  \begin{array}{cc}
		  	i\sin\alpha &  1\\
			1			& -i\sin\alpha
		  \end{array} 
	\right), \label{E:PTHamiltonian}
\end{align}
from which the coupled-mode equations for an optical coupler can be derived as:
\begin{align}
	i\frac{\partial A}{\partial z} 
	&= i\,s\,\sin\alpha\, A + s\,B,\\
	i\frac{\partial B}{\partial z} 
	&= s\,A - i\,s\,\sin\alpha\, B.	
\end{align}
Here, the coupling strengths $h_{AB}$ and $h_{BA}$ between two channels are equal. There also exists gain and loss in the channels $A$ and $B$ with coefficients $\pm \sin \alpha$, respectively. 
It can be checked that the corresponding eigenvalues of this Hamiltonian are $E_\pm=\pm s\cos\alpha$, where $\alpha$ is introduced as a Hermiticity parameter.  When $\alpha=0$, the Hamiltonian $H$ returns to a Hermitian one. 
When $\alpha \neq 0$, this Hamiltonian is not Hermitian, i.e., $H \neq H^\dag$, but  gives the right eigenstates $|E_\pm^R\rangle$ and left eigenstates $|E_\pm^L\rangle$, respectively.
Here, we define $H|E_\pm^R\rangle=E_\pm|E_\pm^R\rangle$ and $H^\dagger|E_\pm^L\rangle=E_\pm|E_\pm^L\rangle$, which have the following explicit forms:
\begin{eqnarray}
|E_+^R(\alpha)\rangle &=& \frac{e^{i\alpha/2}}{\sqrt{2\cos\alpha}}
\left( \begin{array}{c}
1\\
e^{-i\alpha}
\end{array}\right),\\
|E_-^R(\alpha)\rangle &=& \frac{i e^{-i\alpha/2}}{\sqrt{2\cos\alpha}}
\left(\begin{array}{c}
1\\
-e^{i\alpha}
\end{array}\right),
\end{eqnarray}
and  $|E_\pm^L(\alpha)\rangle=|E_\pm^R(-\alpha)\rangle$. The  eigenvalue equation of $H$ can be written as
\begin{align}
\Xi^{\dagger}H\Phi=\tilde{E},
\end{align}
where $\Phi\equiv (|E_+^R\rangle,|E_-^R\rangle)^T$, $\Xi\equiv (|E_+^L\rangle,|E_-^L\rangle)^T$, and the diagonal matrix $\tilde{E}$ consists of the energy spectrum:
\begin{align}
	\tilde{E}
	=\left(
		\begin{array}{cc}
			E_+	&0\\
			0	&E_-
		\end{array}
	 \right).
\end{align}

It is known that a non-Hermitian matrix does not have an orthogonal set of eigenvectors~\cite{Matrix}.
In other words, a non-Hermitian matrix, in general, cannot be transformed into a diagonal form by an orthogonal matrix. 
Nevertheless, the left and right eigenstates of $H$ have the {\it biorthogonality} property: 
\begin{eqnarray}
\Xi^{\dagger}\, \Phi=I.\label{cEq3}
\end{eqnarray}
Therefore, the Naimark dilation theorem allows us to embed the original $2\times 2$ non-Hermitian Hamiltonian $H$ into a $4\times 4$ Hermitian one:
\begin{align}\label{eq_11}
	\mathbf{H}
	=
	\frac{\cos\alpha}{2}
	\begin{pmatrix}
		H\eta^{-1} + \eta H		& H-H^\dagger\\
		H^\dagger-H				& H\eta^{-1} + \eta H
	\end{pmatrix},
\end{align}
where $\eta\equiv\Xi\Xi^\dagger$ is the metric operator coming from the Naimark dilation in Ref.~\cite{Naimark}.  
To answer the question of whether a unitary transformation exists that can rotate the $\mathcal{PT}$-symmetric system shown in Eq. (\ref{E:PTHamiltonian}) to one with only real values, note the structure of our dilated Hamiltonian $\mathbf{H}$ shown in Eq.~(\ref{eq_11}).
Since $\mathbf{H}$ is Hermitian, it is simple to use the {\it unitary transformation} $\mathbf{U}$ to change the basis and find the different mathematical representations of the physics system $\mathbf{H}$. It is important to note that the connection between $H$ and $\mathbf{H}$ is based on the constraint of the dilated quantum state $\mathbf{\Psi}$ having the form of
\begin{align}
	\mathbf{\Psi}
	\propto
	\begin{pmatrix}
		\psi\\
		\eta\psi
	\end{pmatrix}, \label{E:dilatedState}
\end{align}
and one can see the relation by applying $\mathbf{H}$ on $\mathbf{\Psi}$ and obtains that
\begin{align}
	\mathbf{H}\mathbf{\Psi} 
	= 
	\begin{pmatrix}
		H	&0\\
		0	&H^\dagger	
	\end{pmatrix}
	\begin{pmatrix}
		\psi\\
		\eta\psi
	\end{pmatrix}. \label{E:connectPTandHrmtn}
\end{align}
To preserve the structure in Eqs. (\ref{E:dilatedState}-\ref{E:connectPTandHrmtn}), the unitary transformation to be applied onto $\mathbf{H}$ is
\begin{align}
	\mathbf{U}
	=
	\begin{pmatrix}
		U	&0\\
		0	&U^\dagger
	\end{pmatrix},
\end{align}
where $U$ lives in the two dimensional Hilbert space and gives a new Hamiltonian $H^\prime=UHU^{\dagger}$; while $H'$ keeps the same spectrum as $H$. The corresponding left and right eigenstates of $H^\prime$ are $\Xi^\prime=U\Xi$ and $\Phi^\prime =U\Phi$, respectively.
With the help of the Bloch sphere, we can regard this unitary transformation as a rotation operator by decomposing the operator $U$ with the Euler angles $\phi_z$, $\phi_{y\prime}$ and $\phi_{z^\prime}$, i.e.,
\begin{align}\label{eq_U}
	U = e^{-i\sigma_z\phi_z/2} e^{-i\sigma_y\phi_{y'}/2} e^{-i\sigma_z\phi_{z'}/2},
\end{align}
where $z$ corresponds to the rotation axis; while $y^\prime$ and $z^\prime$ indicate the new axis after rotation. Here, $\sigma_z$ and $\sigma_y$ are the corresponding Pauli matrices; $\phi_{z}$, $\phi_{y'}$, and $\phi_{z'}$ are the corresponding Euler angles.
Then, we can derive an explicit form for the generalized $H'$ with the following matrix elements:
\begin{align*}
	&h'_A 
	= -(\cos\phi_{z'} \sin\phi_{y'} - i\sin\alpha \cos\phi_{y'}),\\
	&h'_{AB} 
	=  e^{-i\phi_z} (\cos\phi_{z'}\cos\phi_{y'} - i\sin\phi_{z'} + i\sin\alpha \sin\phi_{y'}),\\
	&h'_{BA}
	= e^{i\phi_z}(\cos\phi_{z'}\cos\phi_{y'} + i\sin\phi_{z'}+i\sin\alpha\sin\phi_{y'}),\\
	&h'_B
	= \cos\phi_{z'} \sin\phi_{y'} - i\sin\alpha \cos\phi_{y'},
\end{align*}
and compute its eigenvalues:
\begin{align*}
E_\pm=\frac{(h'_A+h'_B)\pm\sqrt{(h'_A+h'_{B})^2-4(h'_A h'_{B}-h'_{AB} h'_{BA})}}{2}.
\end{align*}

With this unitary transformation, the condition of the $\mathcal{PT}$-symmetry in the new Hamiltonian $H'$ is changed to $\mathcal{P}^\prime \mathcal{T}^\prime H^\prime\mathcal{P}^\prime \mathcal{T}^\prime=H'$, where $\mathcal{P}^\prime$ and $\mathcal{T}^\prime$ are the new parity and time reversal operators after the unitary transformation.
It is worth remarking that $\mathcal{T}^\prime$ is equal to a unitary operator multiplied by an anti-linear operator. It is the general form of a time reversal operator.
Both $\mathcal{P}^\prime$ and $\mathcal{T}^\prime$ satisfy the conditions of  $\mathcal{P}^{\prime 2}=\mathcal{T}^{\prime 2}=\mathbf{I}$ and $[\mathcal{P}^\prime,\mathcal{T}^\prime]=0$. 

In general, there are eight degrees of freedom in an arbitrary Hamiltonian. Among them, four degrees of freedom correspond to three Euler angles introduced in Eq.~(\ref{eq_U}) and the non-Hermiticity $\alpha$.   
We can also introduce the following three constraints for the energy spectrum $E_\pm$ to be satisfied: (1): $E_++E_-=0$, (2): $E_+-E_-=\omega_0$, and (3): $E_+$, $E_-\in\mathcal{R}$.
Here, the first constraint comes from the fact that the overall energy shift should not affect the physical phenomena. The second constraint helps us to focus on physical systems with the same energy scale, denoted by $\omega_0$; while the third ensures all the energies are real values. Finally, we can derive  $\mathcal{PT}$-symmetric Hamiltonians coping with all these constraints.

In addition to the original  $\mathcal{PT}$-symmetric Hamiltonian in Eq.~(\ref{E:PTHamiltonian}), we explicitly list all other possible $\mathcal{PT}$-symmetric Hamiltonians for $2\times2$ couplers:
\begin{subequations}\label{E:generalH}
\begin{align}
	H_1 
	&= s
	\left(
		\begin{array}{cc}
			0 				& 1-\sin\alpha \\
			1 + \sin\alpha	& 0
		\end{array}
	\right),	\label{E:generalH1}\\
	H_2 
	&= s
	\left(
		\begin{array}{cc}
			i\sin\alpha	&-i \\
			i 				& -i\sin\alpha
		\end{array}
	\right),	\label{E:generalH2}\\
	H_3
	&= s
	\left(
		\begin{array}{cc}
			0				& -i + i\sin\alpha \\
			i + i\sin\alpha	& 0
		\end{array}
	\right),	\label{E:generalH3}\\
	H_4
	&= s
	\left(
		\begin{array}{cc}
			1				& -i\sin\alpha\\
			-i\sin\alpha	& -1
		\end{array}
	\right),	\label{E:generalH4}\\
	H_5
	&= s
	\left(
		\begin{array}{cc}
			1			& -\sin\alpha \\
			\sin\alpha	& -1
		\end{array}
	\right) \label{E:generalH5}.
\end{align}
\end{subequations}
Note that the Hamiltonians $H_1$ and $H_5$ shown in Eqs. (\ref{E:generalH1}) and (\ref{E:generalH5}) contain only  real numbers matrix elements.  In terms of the interacting optical channels, the non-Hermiticity in the $\mathcal{PT}$-symmetric system comes from the asymmetric couplings between two channels.
Since there is no complex number involved, the gain and loss effects can be absent to satisfy the $\mathcal{PT}$-symmetry condition.

To realize these passive $\mathcal{PT}$-symmetric couplers as optical devices, one may need to implement asymmetric couplings between two channels.
Physically, it is the difference between gain and loss that contributes to asymmetric coupling.
Take the Hamiltonian $H_1$ shown in Eq. (\ref{E:generalH1}) as an example, as illustrated in Fig. 1(b), asymmetric couplers  in the slab waveguides, dissimilar optical fibers, or coupled cavities with chiral mirrors~\cite{chiral} are ready to act as  $\mathcal{PT}$-symmetry-based functional devices.

We remark that, although a seemingly  general $\mathcal{PT}$-symmetric Hamiltonian below was proposed  in~\cite{bender2007}:
\begin{align}
	H
	=
	\left( 
		\begin{array}{cc}
			x+(z+iy) &\frac{z}{\tan\gamma}-iy\tan\gamma\\
			\frac{z}{\tan\gamma}-iy\tan\gamma &x-(z+iy)
		\end{array}
	\right), \label{E:BenderPTHamiltonian}
\end{align}
with $x,y,z,\gamma\in\mathcal{R}$, under the condition of fixed energy difference, it is unitarily equivalent to $H'$ with $\phi_{z'}=\phi_{z}=0$. The rotation operator used to generate $H'$ gives an explicit picture that Eq.~(\ref{E:BenderPTHamiltonian}) can be obtained by only rotating Eq. (\ref{E:PTHamiltonian}) about $y$-axis, which is obviously not the most general $2 \times 2$ $\mathcal{PT}$-symmetric Hamiltonian. 

Another feature which is not obvious in our generalized Hamiltonian $H'$ and Eq.~(\ref{E:generalH1}-\ref{E:generalH5}) is the symmetry broken phase where the eigenvalues become complex. The absence of this feature is due to the fact that the range of $\sin\alpha$ in Eq.~(\ref{E:PTHamiltonian}) is restricted in $[-1,1]$. To recover the complex eigenvalue feature, one can replace $\sin\alpha$ by $\kappa$ where $\kappa\in(-\infty,\infty)$. With the replacement, e.g., Eq.~(\ref{E:generalH1}) becomes
\begin{align}
	\tilde{H}_1 = s
	\left(
		\begin{array}{cc}
			0 				& 1-\kappa \\
			1 + \kappa	& 0
		\end{array}
	\right),
\end{align}
with eigenvalues $\pm\sqrt{1-\kappa^2}$, which are complex when $\kappa>1$ and while all the entries are real.

The unitary matrix providing the generalization here is not as trivial as the basis transformation in conventional quantum mechanics theory. In $\mathcal{PT}$-symmetric theory, a new inner product is defined so is the meaning of unitary transformation. Thus the matrix of transformation in this paper is unitary in conventional inner product but not in $\mathcal{PT}$-symmetry theory and cannot be consider a basis transformation in $\mathcal{PT}$-symmetry theory. The relation established here firstly illustrates the reason that unitary matrix generating more general Hamiltonian by using Naimark dilation, and then provides a clear picture by decomposing the transformation matrix into rotation operators on Bloch sphere. The idea here does not only apply on two-level system but may also applies on higher dimension, and gives a systematic way to find out other general $\mathcal{PT}$-symmetric Hamiltonian.

In conclusion, based on the Naimark dilation to construct a Hermitian Hamiltonian with the left and right eigenstates from a $\mathcal{PT}$-symmetric system, we reveal a generalized $2\times2$ $\mathcal{PT}$-symmetric coupler without any complex optical potentials involved.  Instead of gain/loss balanced $\mathcal{PT}$-symmetric optical couplers, now passive devices with asymmetric coupling coefficients can also be used  to implement these $\mathcal{PT}$-symmetric optical systems.

This work is supported in part by the Ministry of Science and Technologies, Taiwan, under the contract No. 101-2628-M-007-003-MY4. MHH is supported by an ARC Future Fellowship under Grant FT140100574.


\begin{thebibliography}{99}
\bibitem{bender1998} C. M. Bender and S. Boettcher, ``Real spectra in non-Hermitian
Hamiltonians having $\mathcal{PT}$ symmetry," Phys. Rev. Lett. \textbf{80}, 5243 (1998).

\bibitem{bender2007}
C. M. Bender, D. C. Brody, H. F. Jones, and B. K. Meister, 
``Faster than Hermitian quantum mechanics,''
Phys. Rev. Lett. \textbf{98}, 040403 (2007).

\bibitem{YiChan} Y.-C. Lee, M.-H. Hsieh, S. T. Flammia, and R.-K. Lee, ``Local $\mathcal{PT}$ symmetry violates the no-signaling principle,'' Phys. Rev. Lett. 112, 130404 (2014).

\bibitem{PT-device} L. Chang, X. Jiang, S. Hua, C. Yang, J. Wen, L. Jiang,
G. Li, G. Wang, and M. Xiao,
``Parity–time symmetry and variable optical isolation in active–passive-coupled microresonators,''
Nature Photon. \textbf{8}, 524, (2014).

\bibitem{klaiman2008}
S. Klaiman, U. G\"{u}nther, and N. Moiseyev, 
``Visualization of branch points in $\mathcal{PT}$-symmetric waveguides," Phys. Rev. Lett.
\textbf{101}, 080402 (2008).



\bibitem{soliton}
R. Driben and B. A. Malomed, 
``Stability of solitons in parity-time-symmetric couplers," Opt. Lett. \textbf{36}, 4323 (2011).

\bibitem{makris2008}
K. G. Makris, R. El-Ganainy, D. N. Christodoulides, and Z. H. Musslimani, 
``Beam dynamics in $\mathcal{PT}$ symmetric optical lattices,'' Phys. Rev. Lett. \textbf{100}, 103904 (2008).

\bibitem{nature}
A. Regensburger, C. Bersch, M. A Miri, G. Onishchukov, D. N. Christodoulides, and U. Peschel, "Parity-time synthetic photonic lattices," Nature \textbf{488}, 167 (2012).


\bibitem{loss} B. Peng, S. K. Ozdemir, S. Rotter, H. Yilmaz, M. Liertzer, F. Monifi, C. M. Bender, F. Nori, and L. Yang,
``Loss-induced suppression and revival of lasing,'' Science \textbf{346}, 328 (2014).

\bibitem{Guo}
A. Guo, G. J. Salamo, D. Duchesne, R. Morandotti, M. Volatier-Ravat, V. Aimez, G. A. Siviloglou, and D. N. Christodoulides, 
``Observation of $\mathcal{PT}$-symmetry breaking in complex optical potentials," Phys. Rev.
Lett. \textbf{103}, 093902 (2009).

\bibitem{Ruter}
C. E. R\"{u}ter, K. G. Makris, R. El-Ganainy, D. N. Christodoulides, M. Segev, and D. Kip, 
``Observation of parity-time symmetry in optics," Nature Phys. \textbf{6}, 192 (2010).

 \bibitem{passive} 
L. Feng, Y.-L. Xu, W. S. Fegadolli, M.-H. Lu, J. E. B. Oliveira, V. R. Almeida, Y.-F. Chen, and  A. Scherer, ``Experimental demonstration of a unidirectional reflectionless parity-time metamaterial at optical frequencies," Nature Mat.  \textbf{12}, 108 (2013).
 
 \bibitem{super} M.-A. Miri, M. Heinrich, R. El-Ganainy, and D. N. Christodoulides,
``Supersymmetric optical structures,'' Phys. Rev. Lett. \textbf{110}, 233902 (2013).
 
\bibitem{Naimark} 
U. Gunther and B. F. Samsonov, 
``Naimark-dilated $\mathcal{PT}$-symmetric brachistochrone,'' Phys. Rev. Lett. \textbf{101}, 230404 (2008).

\bibitem{Marcuse}
D. Marcuse, {\it ``Theory of dielectric optical waveguides,''} (Academic Press, New York, 1972).

\bibitem{Marcuse-85}
D. Marcuse, ``Directional-coupler filter using dissimilar optical fibres,`` Electron. Lett \textbf{21}, 726 (1985).

\bibitem{Snyder}
A. W. Snyder and J. D. Love, {\it ``Optical waveguide theory,''} Chapman \& Hall, London, 1983).

\bibitem{Plum-2011}
E. Plum, V. A. Fedotov, and N. I. Zheludev, ``Asymmetric transmission of electromagnetic radiation where you do not expect it: when passing through a two-dimensional array of highly symmetric particles,'' \textit{J. Opt.,} \textbf{13}, 024006 (2011).

\bibitem{Lawerance-2014}
M. Lawrence, N. Xu, X. Zhang, L. Cong, J. Han, W. Zhang, and S. Zhang, ``Manifestation of $PT$ Symmetry Breaking in Polarization Space with Terahertz Metasurfaces,'' Phys. Rev. Lett. \textbf{113}, 093901 (2014).

\bibitem{Matrix} R. A. Horn and C. R. Johnson, {\it ``Matrix analysis,''} (Cambridge University Press, 1985).

\bibitem{chiral} R. B. B. Santos,``Non-Hermitian model for resonant cavities coupled by a chiral mirror,'' EuroPhys. Lett. \textbf{100}, 24005 (2012).

\end{thebibliography}
\end{document}